# Collimated GeV attosecond electron-positron bunches from a plasma channel driven by 10 PW lasers

**Authors:**

Xing-Long Zhu[1, 4], Zheng-Ming Sheng[1, 2, 4, 5†], Min Chen[1, 4], Tong-Pu Yu[3], and Su-Ming Weng[1, 4]

**Affiliations:**

[1] Key Laboratory for Laser Plasmas (MoE), School of Physics and Astronomy, Shanghai Jiao Tong University, Shanghai 200240, China

[2] SUPA, Department of Physics, University of Strathclyde, Glasgow G4 0NG, UK

[3] Department of Physics, National University of Defense Technology, Changsha 410073, China

[4] Collaborative Innovation Center of IFSA, Shanghai Jiao Tong University, Shanghai 200240, China

[5] Tsung-Dao Lee Institute, Shanghai 200240, China

[†]e-mail: zmsheng@sjtu.edu.cn

**Abstract**

**High-energy positrons and bright γ-ray sources are unique both for fundamental research and practical applications. However, GeV electron-positron pair jets and γ-ray flashes are still hardly produced in laboratories. Here we demonstrate that, by irradiating two 10 PW-scale laser pulses onto a near-critical density plasma channel, highly-directional GeV electron-positron pairs and bright γ-ray beams can be efficiently generated. Three-dimensional particle-in-cell simulations show that GeV positron jets show high density ($8 \times 10^{21}$/cm$^3$), attosecond duration (400 as) and a divergence angle of 14°. Additionally, ultrabright ($2 \times 10^{25}$ photons/s/mm²/mrad²/0.1%BW) collimated attosecond (370 as) γ-ray flashes with a laser energy conversion efficiency of 5.6% are emitted. Once realized in experiment, it may open up new possibilities for a wide variety of applications.**

**Introduction**

Since Anderson observed positrons [1], researches on positrons have excited great interest and play an important role in various domains [2,3], including fundamental science, medicine, and industry. Compared with the conventional positron sources, laser-driven positron sources have many potential advantages, such as having a high energy, yield and density, ultrashort beam size, etc. Currently, by use of high power intense lasers, multi-MeV positrons can be easily produced in laboratories [4-7]. However, giant highly-energetic (i.e., GeV and TeV energies) positron jets with extremely high-density are still out of reach, which occur only in energetic astrophysical environments [2,8,9], such as γ-ray bursts, pulsars and black holes. It is very difficult to achieve such positron sources on earth with the current laser technologies or traditional methods. On the other hand, the atto-beams of relativistic electrons and X/γ-rays [10-13] show powerful tools for diverse scientific research and technical applications, enabling the time-space imaging with sub-atomic resolution in attosecond regime, however, the atto-beam property of laser-driven positrons has been scarcely investigated.

Several ongoing and arranged laser facilities [14-17] will deliver ultrahigh intensity laser pulses of $10^{23-24} \text{W/cm}^2$ and power of 10-200 PW. This would open a new realm of possibilities for light-matter interactions in the radiation and quantum-dominated regime [18,19]. The proposed schemes show that when the laser intensity is above $10^{23} \text{W/cm}^2$, high-energy dense positron sources can be produced significantly via the multiphoton Breit-Wheeler (BW) process [20] from various media such as plasmas [21-27] or relativistic electron beams [28-30]. However, so far, the production of dense positron jets and bright γ-ray flashes with both atto-scale beam duration and GeV energies at currently available laser systems has not yet been achieved.

In this paper, we present a practical approach to generate collimated GeV positron beams and bright γ-ray flashes with attosecond duration at an achievable laser intensity of $\sim 10^{22} \text{W/cm}^2$. Figure 1(a) shows the sketch of our scenario. A 10 PW drive laser pulse irradiates a near-critical-density (NCD) plasma channel from the left side and propagates along the *x*-axis direction [see Fig1(b)], where the channel density profile is radially symmetric. During the laser propagation, the pulse intensity is greatly enhanced in the plasma channel. Meanwhile, dense attosecond electrons are trapped by the intense pulse and accelerated to multi-GeV energies, so that energetic attosecond γ-rays are efficiently emitted via nonlinear Compton scattering (NCS) [13,31]. The dense GeV atto-beams of electrons and γ-rays then collide with the second probe pulse from the right side, and the multiphoton BW process is triggered, resulting in abundant dense GeV positrons with atto-scale beam duration. As a comparison, we also consider the plasma with a uniform density distribution, as shown in Fig. 1(c).

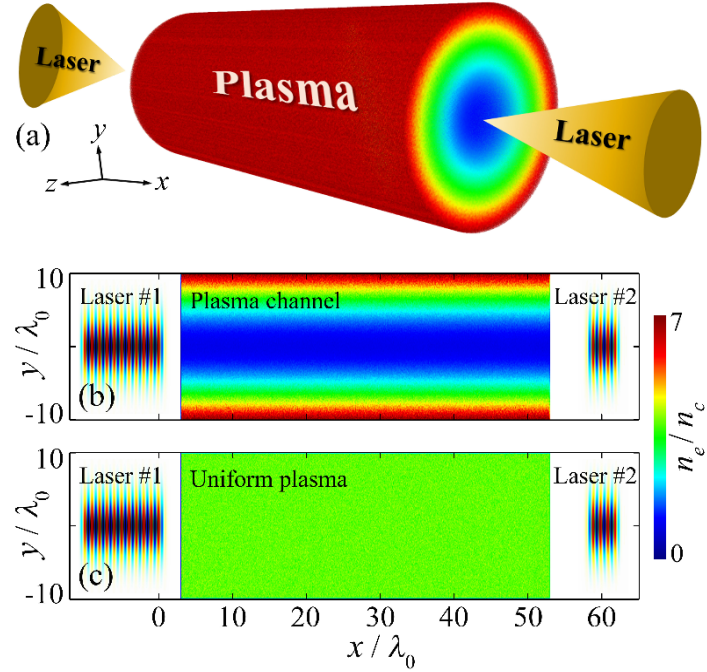

**Fig. 1.** (**a**) Schematic of collimated GeV positron jets generation by two 10 PW laser pulses interaction with a plasma channel filled-in with NCD plasmas. The drive pulse (laser #1) and probe pulse (laser #2) are incident from the left and right sides, respectively. Two cases are considered: the radially symmetric density profile in (**b**) and uniform density profile in (**c**).

**Results**

Three-dimensional (3D) particle-in-cell (PIC) simulations were performed using the code EPOCH [32] with both QED and collective plasma effects incorporated [33,34]. In the simulations, two 10 PW-scale high-power linearly-polarized Gaussian laser pulses (drive laser and probe laser) are incident with a time delay of $55T_0$ from the left side and right side of the box, respectively. The temporal profile of both laser pulses are trapezoidal with a duration of $12T_0$ $(1-10-1T_0)$ for the drive pulse and $5T_0$ $(1-3-1T_0)$ for the probe pulse. The normalized amplitude of both lasers is $a_0 = eE_0/m_e c\omega_0 = 150$, corresponding to a currently approachable intensity of $3\times 10^{22} \text{W/cm}^2$ in laboratory [35], with a focus spot of $\sigma_0 = 4\lambda_0$. Here $e$ is the unit charge, $m_e$ is the electron mass, $\omega_0$ is the laser oscillation frequency, $\lambda_0 = T_0 c =$ $1\mu m$ is the laser wavelength, and $c$ is the speed of light in vacuum. The NCD plasma has a transverse density profile of $n_e = n_0 + \Delta n(r^2/\sigma_0^2)$ in the plasma channel located between 3 and $53\lambda_0$, where $n_c = m_e\omega_0^2/4\pi e^2$ is the critical density, $n_0 = 1n_c$, $\Delta n = 0.1a_0 n_0/\sigma_0^2(\mu m^2)$, and $r = y^2 + z^2$ is the radial distance from the channel axis. For the case with the uniform plasma, the density is $n_e = 3.9n_c$ to keep the total number of plasma electrons unchanged. The simulation box is $x \times y \times z = 60 \times 20 \times 20\lambda_0^3$ with a cell of $\Delta x \times \Delta y \times \Delta z = \lambda_0/30 \times \lambda_0/12 \times \lambda_0/12$ and 16 macro-particles in each cell. In order to save the computing sources, a moving window is employed in all simulations below.

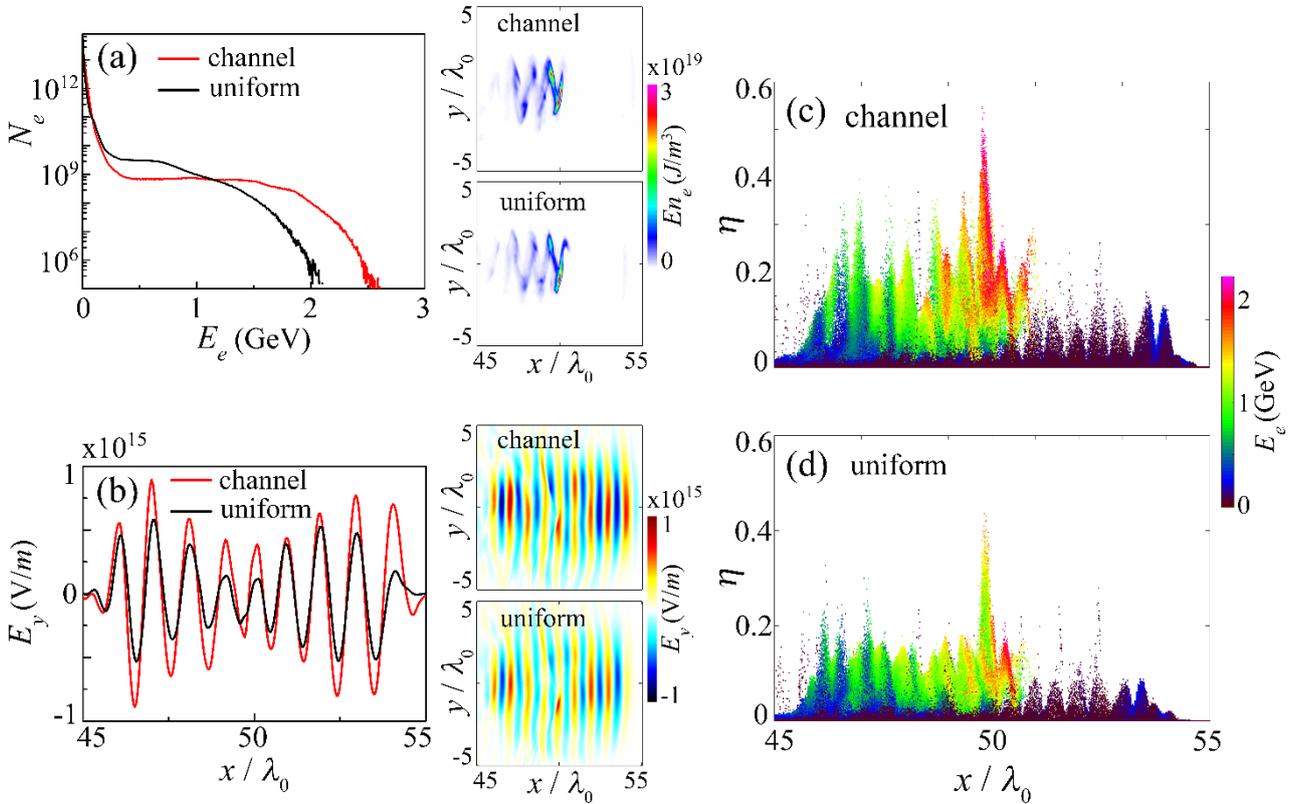

**Fig. 2.** (**a**) The energy spectra of electrons. (**b**) The transverse electric fields along the laser propagation axis. The insets show the distributions of the trapped electron energy density (**a**) and laser transverse electric field (**b**), respectively, in the case with a plasma channel (top) and a uniform plasma (bottom). Distribution of the parameter $\eta$ along the *x*-axis in the case with a plasma channel (**c**) and a uniform plasma (**d**).

Figure 2(a) illustrates the distributions of the electron energy density and energy spectrum. It is shown that the electrons in the plasma channel can be accelerated to much higher energy than that in uniform plasma case. This can be attributed to the coupling effects of high-intensity laser interaction with NCD plasmas [36,37], where the plasma channel works as a optical lens to enhance the intensity of laser pulse significantly, as presented in Fig. 2(b). Since a large number of electrons are confined in the high intensity area of the laser pulse, they consume the most laser energy by emitting high-energy γ-rays during the rapid acceleration. It is interesting to note that the laser intensity is still enhanced by four times within the plasma channel. The beam energy density of the accelerated electrons is as high as $3 \times 10^{19}$ J/m$^3$ with a high energy of 2.5 GeV and a ultrashort beam duration of several hundreds of attoseconds, which is more than eight-fold higher than the threshold of high-energy-density physics (HEDP) [38]. Such energetic electrons interacting with the extremely-intense laser fields results in $\eta > 0.1$, where $\eta = (\gamma_e/E_s)|\mathbf{E}_\perp + \mathbf{v} \times \mathbf{B}|$ is the critical parameter for determining the importance of quantum-dominated radiation emission [26,34,39], as shown in Figs. 2(c) and 2(d). Here, $\gamma_e$ is the electron Lorentz factor, $\mathbf{E}_\perp$ is the electric field perpendicular to the electron velocity $\mathbf{v}$, and $E_s = m_e^2 c^3/e\hbar$ is the Schwinger field. Finally, most of the drive laser pulse energy is absorbed by the plasma and the electrons are accelerated to GeV energies, together with bright γ-ray emission. Such dense GeV atto-beams may open new possibilities for a number of applications, providing ultrahigh time-resolved studies with attosecond-scale resolution in various scientific frontiers such as high energy physics, plasma physics, astrophysics, and so on.

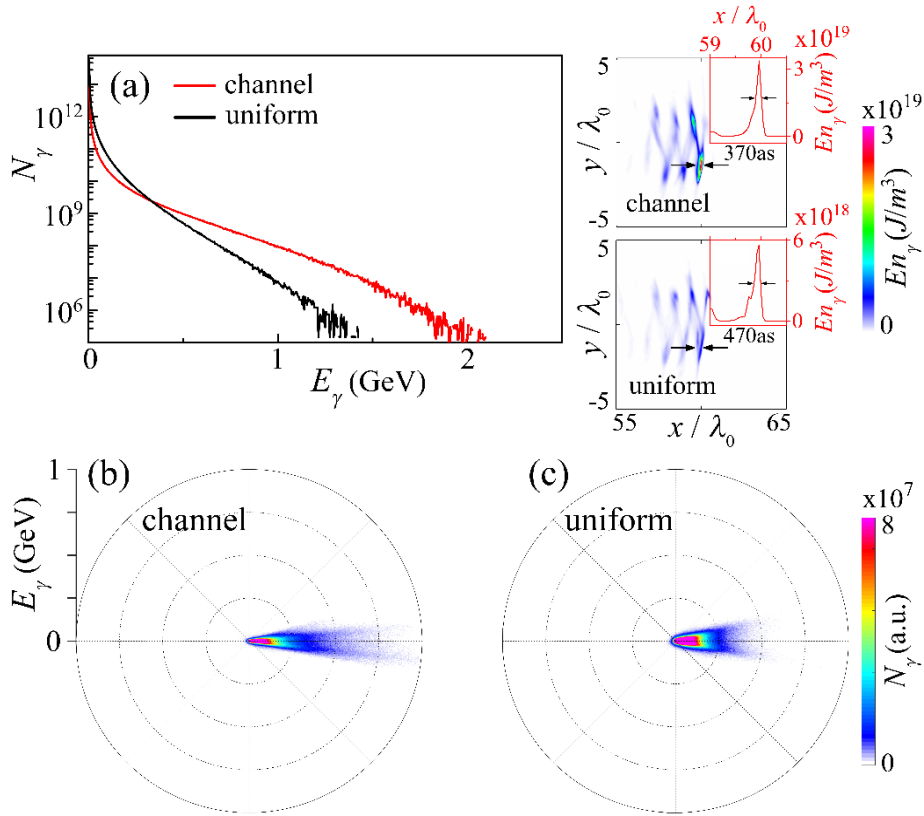

**Fig. 3.** (a) The energy spectra of γ-rays. The insets show the energy density distribution of the γ-ray emission with a plasma channel (top) and a uniform plasma (bottom), respectively. The angular energy distribution of the emitted γ-rays in the plasma channel (**b**) and uniform plasma (**c**).

Since the critical parameter $\eta$ in the case with plasma channel is much larger than that with the uniform plasma, GeV γ-rays are efficiently emitted with a high photon energy density, as shown in Fig. 3(a). Here, the attosecond γ-rays radiated have a smaller divergence angle with high photon energies than that with the uniform plasma, as presented in Figs. 3(b) and 3(c). With the plasma channel, the γ-ray beam has a total photon yield of $2.5 \times 10^{11}$ at 25 MeV, a full width at half maximum (FWHM) cross-section of ~1.5μm$^2$, a FWHM divergence of $0.1 \times 0.1 \text{rad}^2$, and a total pulse duration of ~900as at FWHM. The results indicate that the GeV γ-ray with a peak brightness of ~$2 \times 10^{25}$ photons/s/mm$^2$/mrad$^2$/0.1%BW is obtained, which is several orders of magnitude higher than the presented in current laboratories [40-43] and is also much brighter than the level in other simulations [26,44-46]. Meanwhile, the γ-ray beam is characterized with a desirable ultrashort duration of <370as per pulse, which may open the door to a new realm of ultrafast-science research in a wide range of scientific fields.

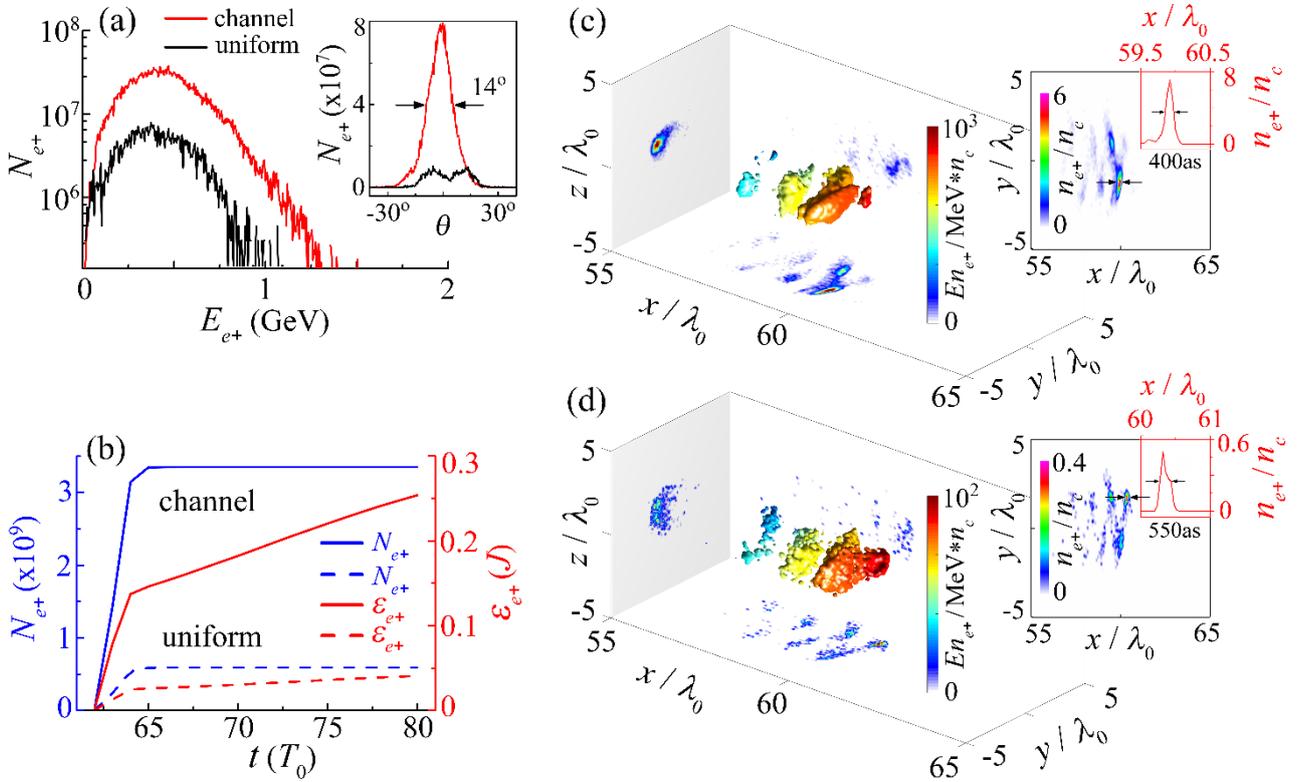

**Fig. 4.** (**a**) The energy spectrum of positrons with a plasma channel (red line) and a uniform plasma (black line), where the inset shows the positron angular divergence. (**b**) Evolution of the total yield and energy of positrons with the plasma channel (top) and a uniform plasma (bottom), respectively. The angular energy distribution of the emitted γ-rays in the plasma channel (**b**) and uniform plasma (**c**).

Since the critical parameter $\eta$ in the case with plasma channel is much larger than that with the uniform plasma, GeV γ-rays are efficiently emitted with a high photon energy density, as shown in Fig. 3(a). Here, the attosecond γ-rays radiated have a smaller divergence angle with high photon energies than that with the uniform plasma, as presented in Figs. 3(b) and 3(c). With the plasma channel, the γ-ray beam has a total photon yield of $2.5 \times 10^{11}$ at 25 MeV, a full width at half maximum (FWHM) cross-section of ~1.5μm$^2$, a FWHM divergence of $0.1 \times 0.1 \text{rad}^2$, and a total pulse duration of ~900as at FWHM. The results indicate that the GeV γ-ray with a peak brightness of ~$2 \times 10^{25}$ photons/s/mm$^2$/mrad$^2$/0.1%BW is obtained, which is several orders of magnitude higher than the presented in current laboratories [40-43] and is also much brighter than the level in other simulations [26,44-46]. Meanwhile, the γ-ray beam is characterized with a desirable ultrashort duration of <370as per pulse, which may open the door to a new realm of ultrafast-science research in a wide range of scientific fields.

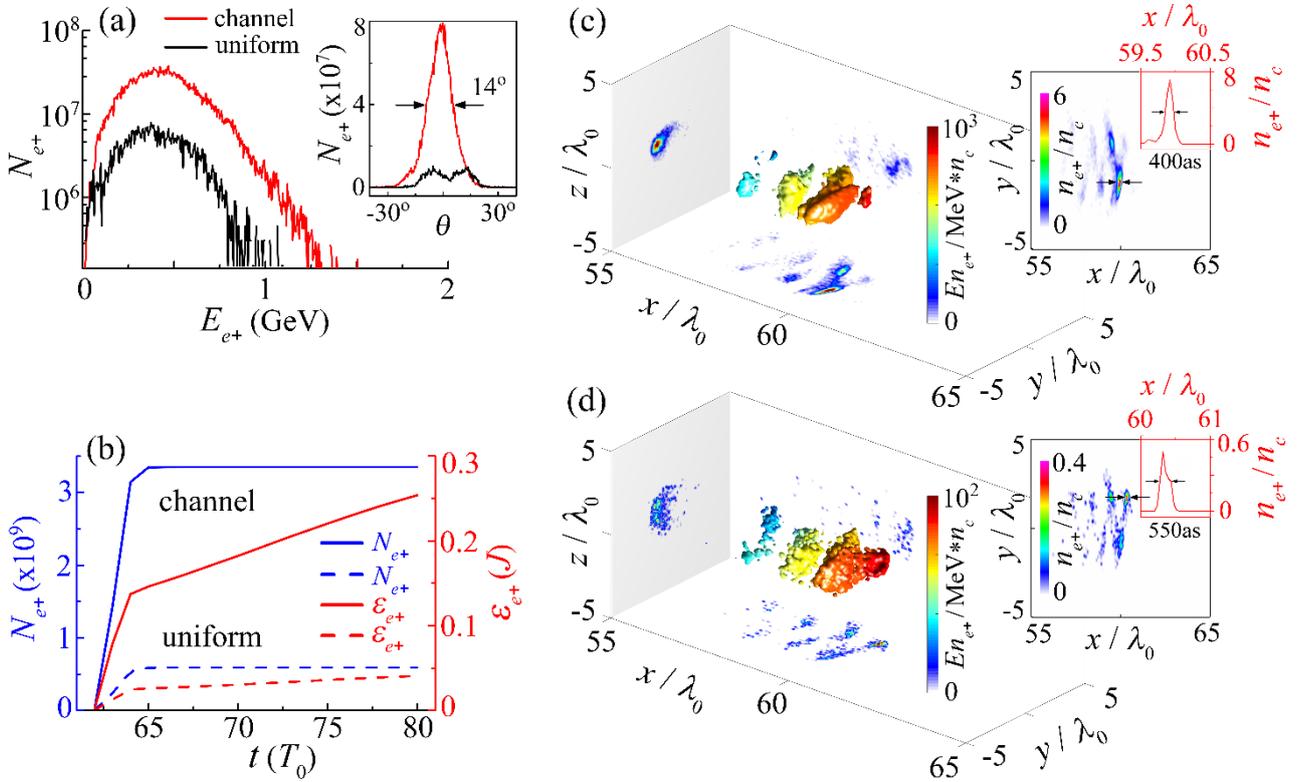

**Fig. 4.** (**a**) The energy spectrum of positrons with a plasma channel (red line) and a uniform plasma (black line), where the inset shows the positron angular divergence. (**b**) Evolution of the total yield and energy of positrons with the interaction time. The energy density distribution of positrons generated from a plasma channel (**c**) and a uniform plasma (**d**). The insets present the density profile of positrons in the *x-y* cross section.

The extremely-strengthened laser pulse accelerates the electrons to multi-GeV scale. During this process, giant energetic γ-ray flashes are emitted via NCS in the NCD plasmas. Both of such accelerated electrons and emitted γ-photons with GeV energies then collide with the probe laser pulse incident from the right side, triggering the multiphoton BW process. This can be described by the quantum parameter $\chi = (\hbar\omega/2m_e c^2)|\mathbf{E}_\perp + \hat{\mathbf{k}} \times c\mathbf{B}|/E_s$ [26,34,39], where $\hbar\omega$ and $\hat{\mathbf{k}}$ are the energy and unit vector of the emitted photons. As a result, high-yield, well-collimated GeV positron jets are effectively generated with an overdense density profile and attosecond-scale beam duration. Figure 4 shows the 3D PIC simulation results of positrons generation in both cases above. One can see clearly that the jets can be significantly enhanced in the case with plasma channel. The collimated GeV positron jets obtained have a high density of $7n_c$ with a small divergence angle of 14° and a ultrashort beam duration of 400as at FWHM. The energy density of such positrons is about $10^{18}$J/m$^3$, which is $10^7$-fold higher than the HEDP threshold. This would offer exciting new tools for positron-based researches, and potentially pushing such studies into the attosecond regime.

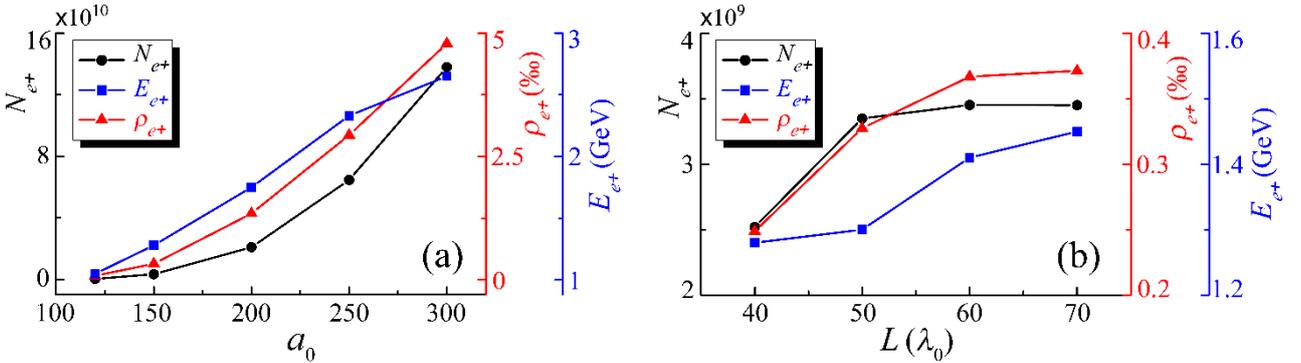

**Fig. 5.** The yield $N_{e^+}$, energy conversion efficiency $\rho_{e^+}$, and cutoff energy $E_{e^+}$ of the positrons as a function of the laser amplitude $a_0$ (**a**) and the plasma longitudinal length $L$ (**b**).

**Discussion and conclusion**

The present simulations demonstrate a promising and efficient approach for generating well-collimated, energetic, dense positron jets with attosecond-scale beam duration from a NCD plasma channel, which is driven by 10 PW-scale intense lasers. In order to further explore the parametric effects and robustness of this scheme, a series of 3D PIC simulations are carried out by employing different NCD plasma channels and laser intensities. First, the effect of laser intensity on the jet generation is investigated, where all other parameters are kept the same as before except for the laser amplitude of $a_0$ and the corresponding plasma density of $n_e$. Figure 5(a) shows the simulation results. One can see that as the laser intensity increases, the efficiency of laser-produced positrons is enhanced significantly. This is due to the fact that, with the increase of laser intensity, the multiphoton BW process can be easily triggered. Note that the critical quantum parameter $\chi \sim \hbar\omega|\mathbf{E}_\perp|/m_e c^2 E_s \propto \hbar\omega a_0$. With the forthcoming multi-PW laser facilities, our scheme potentially gives rise to highly-efficient dense GeV positron jets and bright γ-ray flashes with desirable attosecond-scale beam property.

Figure 5(b) shows that the simulation results with different plasma channel length $L$ ranging from $40\lambda_0$ to $70\lambda_0$. It is shown that a longer channel length is very useful for the production of positrons. Both the maximum energy and yield of positrons obtained increase with $L$, together with a higher energy conversion efficiency from the laser pulse to the positrons. This is attributed to the accelerated electrons, which can obtain a higher energy with a longer acceleration distance and efficiently radiate giant energetic γ-rays in NCD plasmas [36,37]. However, further increase of the channel length is not always better. For example, the generation of positrons saturates for $L > 60\lambda_0$, because the drive laser pulse is rapidly depleted in such a long plasma channel, so that the electron acceleration and γ-ray emission become limited and the electron-positron pair production does not enhance any more. This could be used for tuning and enhancing the positron jet generation in future experiments.

In summary, we have investigated the generation of collimated GeV attosecond positron beams from 10 PW laser interaction with the NCD plasma at a currently achievable laser intensity of $\sim 10^{22} \text{W/cm}^2$. It is shown that high-yield, well-collimated, dense GeV attosecond positron beams are efficiently produced within a plasma channel. Compared with the uniform plasma case, the positron production is greatly enhanced due to the strong focusing and enhancement of the incident laser pulse in the plasma channel. The yield, energy conversion efficiency, and cutoff energy of the positrons obtained increase with the incident laser intensity, which can be further enhanced by using a long plasma channel. With the upcoming next-generation laser facilities (e.g., ELI [14], XCELS [15], Apollon [16], and SULF [17]), such collimated dense GeV positron jets and bright γ-ray flashes both with desirable atto-beam capability may open new avenues for ultrafast studies in physics, chemistry, biology, etc.

**Acknowledgements**

This work was in part supported by the Science and Technology Commission of Shanghai Municipality (Grant No. 16DZ2260200), the Ministry of Science and Technology of China for an International Collaboration Project (Grant No. 2014DFG02330), NSFC (Grant Nos. 11721091, 11774227, 11622547, and 11655002), the National Key Research and Development of China (Grant No. 2018YFA0404802), Hunan Provincial Natural Science Foundation of China (Grant No. 2017JJ1003), and a Leverhulme Trust Grant at the University of Strathclyde. The EPOCH code was partially funded by the UK EPSRC grant EP/G056803/1. The simulations were supported by the PI supercomputer at Shanghai Jiao Tong University.